\documentstyle[manuscript,aps]{revtex}

\begin{document}



\preprint{HEP-95-12}

\title{Quasi-Exactly Solvable Systems and Orthogonal Polynomials}

\author{Carl M. Bender}
\address{Department of Physics, Washington University, St. Louis, MO 63130}

\author{Gerald V. Dunne}
\address{Department of Physics, University of Connecticut, Storrs, CT 06269}


\maketitle

\begin{abstract}
This paper shows that there is a correspondence between quasi-exactly solvable
models in quantum mechanics and sets of orthogonal polynomials $\{ P_n\}$. The
quantum-mechanical wave function is the generating function for the $P_n (E)$,
which are polynomials in the energy $E$. The condition of quasi-exact
solvability is reflected in the vanishing of the norm of all polynomials whose
index $n$ exceeds a critical value $J$. The zeros of the critical polynomial
$P_J(E)$ are the quasi-exact energy eigenvalues of the system.
\end{abstract}


In quantum mechanics there exist potentials for which it is possible to find a
finite portion of the energy spectrum and associated eigenfunctions exactly and
in closed form. These systems are said to be quasi-exactly
solvable.\cite{Tur1,Tur2,Mor,Tur3,Ush}
In such systems the potential depends on a parameter $J$; for positive integer
values of $J$ one can find $J$ eigenvalues and eigenfunctions exactly. The
usual approach to the analysis of quasi-exactly solvable systems is an
algebraic one in which the Hamiltonian is expressed as a nonlinear combination
of generators of a Lie algebra, not belonging to the center of the
corresponding enveloping algebra.\cite{Tur3} This technique is a modification
of the dynamical symmetry approach to exactly solvable quantum-mechanical
systems, in which one can find by algebraic means the {\sl entire} spectrum in
closed form.\cite{Iac}

In this paper we propose an alternative approach to quasi-exact solvability. We
show that the solution $\psi$ to the Schr\"odinger equation for a quasi-exactly
solvable model,
\begin{eqnarray}
H \psi = E \psi,
\label{e1}
\end{eqnarray}
is the generating function for a set of polynomials $\{ P_n (E)\}$ in the
energy
variable $E$. These polynomials satisfy a three-term recursion relation and
therefore form an orthogonal set with respect to some weight function $w(E)$.
 For positive integer values of the parameter $J$, corresponding to quasi-exact
solvability, we find that the norm of $P_n(E)$ {\sl vanishes} for $n\geq J$.
Moreover, all polynomials $P_n(E)$ beyond a critical polynomial $P_J(E)$,
factor
into a product of two polynomials, one of which is $P_J(E)$:
\begin{eqnarray}
P_{n+J}(E) = P_J(E) Q_n(E)\quad (n\geq 0).
\label{e2}
\end{eqnarray}
The zeros of the critical polynomial $P_J(E)$ are precisely the quasi-exact
energy eigenvalues of the quantum-mechanical model.

We illustrate these features of quasi-exactly solvable models with the
following
infinite class of Hamiltonians first discussed by A.~Turbiner \cite{Tur1}
\begin{eqnarray}
H = -{d^2\over dx^2} + {(4s-1)(4s-3)\over 4x^2} - (4s+4J-2)x^2 + x^6.
\label{e3}
\end{eqnarray}
Here, $s$ is an arbitrary parameter. When $s$ lies between $1\over 4$ and
$3\over 4$, there is an attractive centrifugal term; for $s$ outside this range
the centrifugal term is repulsive. When $s={1\over 4}$ or $s={3\over 4}$, the
centrifugal core term disappears leaving a nonsingular sextic oscillator
Hamiltonian
\begin{eqnarray}
H = -{d^2\over dx^2} - (4s+4J-2)x^2 + x^6.
\label{e4}
\end{eqnarray}
When the parameter $J$ in Eq.~(\ref{e3}) is a nonnegative integer, the
corresponding Schr\"odinger equation has $J$ exact, closed-form solutions for
any value of $s$.

We seek a solution $\psi(x)$ to the Schr\"odinger equation for $H$ in
Eq.~(\ref{e3}) of the form
\begin{eqnarray}
\psi(x) = {\rm exp}\left (-{1\over 4}x^4\right ) x^{2s-1/2} \sum_{n=0}^{\infty}
\left ( -{1\over 4}\right )^n {P_n(E)\over n!\Gamma(n+2s)}x^{2n}.
\label{e5}
\end{eqnarray}
Observe that when $s={1\over 4}$ this solution becomes an even-parity wave
function of the oscillator Hamiltonian (\ref{e4}); when $s={3\over 4}$,
$\psi(x)$ becomes an odd-parity wave function of $H$ in (\ref{e4}).

Demanding that $\psi(x)$ in Eq.~(\ref{e5}) obey the Schr\"odinger equation
(\ref{e1}) leads to the following recursion relation for the expansion
coefficients $P_n(E)$:
\begin{eqnarray}
P_n(E)=EP_{n-1}(E)+16 (n-1)(n-J-1)(n+2s-2) P_{n-2}(E)\quad (n\geq 2),
\label{e6}
\end{eqnarray}
subject to the initial conditions
\begin{eqnarray}
P_0(E) = 1 \qquad {\rm and} \qquad P_1(E) = E.
\label{e7}
\end{eqnarray}
 From these initial conditions the recursion relation (\ref{e6}) generates a
set
of monic \cite{monic} polynomials, the next four of which are
\begin{eqnarray}
P_2(E)&=&E^2+(32-32J) s,\cr
P_3(E)&=&E^3 + [(160 - 96 J) s - 32 J + 64] E,\cr
P_4(E)&=&E^4+[(448-192J)s-128J+352]E^2+(3072J^2-12288J+9216)s(s+1),\cr
P_5(E)&=&E^5+[(960-320J)s-320J+1120]E^3+[(15360J^2 -81920J + 91136)s^2\cr
&&\quad +(25600J^2-141312J+164864)s+6144J^2-36864J+49152]E.
\label{e8}
\end{eqnarray}

These polynomials have a number of noteworthy properties. First, for all values
of the parameters $s$ and $J$ they form an orthogonal set. This follows from
the
fact that they are generated by a second-order (three-term) recursion
relation.\cite{Bat} The appearance of a three-term recursion relation is a
consequence of the form of the potential in Eq.~(\ref{e3}). For example, the
corresponding recursion relation for an $x^4$ anharmonic oscillator potential,
whose Hamiltonian is not quasi-exactly solvable, is a higher-order recurrence
relation. The
harmonic oscillator system leads to a two-term recursion relation; this system
is exactly solvable rather than quasi-exactly solvable.

Second, from the expansion (\ref{e5}) we can see that the wave function
$\psi(x,E)$ is the generating function for the polynomials $P_n(E)$.

The third and most significant property of the polynomials $P_n(E)$ is that,
when the parameter $J$ takes positive integer values, the polynomials exhibit
the factorization property in Eq.~(\ref{e2}). This factorization occurs because
the third term in the recursion relation (\ref{e6}) vanishes when $n=J+1$, so
that all subsequent polynomials have the common factor $P_J(E)$. This
factorization property holds for all values of the parameter $s$. Furthermore,
this factorization leads to the result that the zeros of the critical
polynomial
$P_J(E)$ are just the quasi-exact energy eigenvalues. This is true because the
expansion in (\ref{e5}) truncates when $E$ is a zero of $P_J(E)$; when this
series truncates the wave function $\psi(x)$ is automatically normalizable.

To illustrate this factorization we list in factored form the first six
polynomials $P_n(E)$ for the case $J=3$:
\begin{eqnarray}
P_0(E)&=&1,\cr
P_1(E)&=&E,\cr
P_2(E)&=&E^2 - 64 s,\cr
P_3(E)&=& E^3 - (128 s + 32) E,\cr
P_4(E)&=& [E^3 - (128 s + 32)E] E,\cr
P_5(E)&=& [E^3 - (128 s + 32)E] (E^2 + 128 s + 192).
\label{e9}
\end{eqnarray}
Observe that $P_3(E)$ is a common factor of $P_n(E)$ for $n\geq 3$. The zeros
of
$P_3(E)$ are
\begin{eqnarray}
E=0,\qquad E=\pm\sqrt{128s+32},
\label{e10}
\end{eqnarray}
which are the three exact energy eigenvalues for the quasi-exactly solvable
Hamiltonian (\ref{e3}) when $J=3$. The corresponding exact eigenfunctions are
obtained by evaluating $\psi(x)$ in Eq.~(\ref{e5}) at these values of $E$:
\begin{eqnarray}
\psi_0(x)&=&{\rm exp}\left (-{1\over 4}x^4\right ) {x^{2s-1/2}\over \Gamma(2s)}
\left (1-{x^4\over 2s+1}\right ),\cr\cr
\psi_+(x)&=&{\rm exp}\left (-{1\over 4}x^4\right ) {x^{2s-1/2}\over \Gamma(2s)}
\left (1-{\sqrt{128s+32}\over 8s}x^2 + {x^4\over 2s}\right ),\cr\cr
\psi_-(x)&=&{\rm exp}\left (-{1\over 4}x^4\right ) {x^{2s-1/2}\over \Gamma(2s)}
\left (1+{\sqrt{128s+32}\over 8s}x^2 + {x^4\over 2s}\right ).
\label{e11}
\end{eqnarray}
Note that the energy levels may be ordered by the number of nodes of the
corresponding wave function.

A fourth property of the polynomials $P_n(E)$ concerns their norms. The norm
(squared) $\gamma_n$ of $P_n(E)$ is defined as an integral:
\begin{eqnarray}
\gamma_n = \int dE\, w(E) [P_n(E)]^2.
\label{e12}
\end{eqnarray}
It is possible to determine the norms of an orthogonal set of polynomials
directly from the recursion relation; it is not necessary to know explicitly
the
weight function $w(E)$ with respect to which the polynomials are
orthogonal.\cite{domain} The
procedure is simply to multiply the recursion relation (\ref{e6}) by
$w(E)E^{n-2}$ and to integrate with respect to $E$. Using the fact that
$P_n(E)$
is orthogonal to $E^k$, $k<n$, we obtain a simple, two-term recursion relation
for $\gamma_n$:
\begin{eqnarray}
\gamma_n=16n(J-n)(2s+n-1)\gamma_{n-1}.
\label{e13}
\end{eqnarray}
The solution to this equation with $\gamma_0=1$ is
\begin{eqnarray}
\gamma_n = {16^n n!\,\Gamma(J)\Gamma(2s+n)\over \Gamma(J-n)\Gamma(2s)}.
\label{e14}
\end{eqnarray}

This equation reveals that the space of orthogonal polynomials arising from
a quasi-exactly solvable model is associated with a nonpositive definite norm.
In particular, we can see from Eq.~(\ref{e14}) that $\gamma_n$ vanishes for
$n\geq J$ if $J$ is a positive integer. The appearance of a vanishing norm
coincides with the factorization mentioned above and is an alternative
characterization of quasi-exact solvability.

It is interesting that while the polynomials $P_{n+J}(E)$ for $n\geq 0$ have
vanishing norm when $J$ is a positive integer, the quotient polynomials
$Q_n(E)$
in Eq.~(\ref{e2}) form a new orthogonal set of polynomials for each value of
$J$.

Having determined the norms $\gamma_n$ of the polynomials $P_n(E)$ it is
natural
to evaluate the integral of the square of the generating function (wave
function) with respect to the weight function $w(E)$:
\begin{eqnarray}
G(x)=\int dE\,w(E) [\psi(x,E)]^2,
\label{e14.1}
\end{eqnarray}
where $\psi(x,E)$ is given in Eq.~(\ref{e5}). Using the orthogonality of the
polynomials $P_n(E)$, we can express $G(x)$ as a confluent hypergeometric
function:
\begin{eqnarray}
G(x)={\Gamma(J)\over\Gamma(2s)} {\rm exp}\left ( -{1\over 2}x^4\right )
\sum_{n=0}^{\infty} {x^{4n+4s-1}\over n!\,\Gamma(n+2s)\Gamma(J-n)}.
\label{e14.2}
\end{eqnarray}
When $J$ is a positive integer, this sum truncates and we find that $G(x)$ can
be expressed as a linear combination of the squares of the $J$ quasi-exact
eigenfunctions of the Hamiltonian $H$ in Eq.~(\ref{e3}). For example, when
$J=3$, we have
\begin{eqnarray}
G(x)&=&{1\over\Gamma(2s)} {\rm exp}\left ( -{1\over 2}x^4\right )x^{4s-1}
\left [ 1 + {x^4\over s} + {x^8 \over 2s (2s+1)} \right ]\cr\cr
&=&\Gamma(2s)\left ({2s+1\over 4s+1}[\psi_0(x)]^2+{s\over 4s+1}[\psi_+(x)]^2
+{s\over 4s+1}[\psi_-(x)]^2\right ),
\label{e14.3}
\end{eqnarray}
where $\psi_0(x)$ and $\psi_\pm (x)$ are taken from Eq.~(\ref{e11}). We
emphasize that this result is highly nontrivial. Expressing $G(x)$ as a linear
combination of the squares of the eigenfunctions requires that one solve an
overdetermined system of $2J-1$ equations for $J$ expansion coefficients.

Let us now investigate the properties of the weight function $w(E)$. From the
polynomials $P_n(E)$ we can calculate the moments of $w(E)$. Let $a_n$
represent the $2n$th moment of $w(E)$:
\begin{eqnarray}
a_n = \int dE\,w(E) E^{2n}.
\label{e15}
\end{eqnarray}
(Because the polynomials have parity symmetry we know that the odd moments
vanish.) We are free to normalize $w(E)$ so that its zeroth moment is unity:
\begin{eqnarray}
a_0 = 1.
\label{e16}
\end{eqnarray}
The remaining moments can then be determined algebraically:
\begin{eqnarray}
a_1&=&32(J-1) s,\cr
a_2&=&32^2(J-1)s(3Js-5s+J-2),\cr
a_3&=&32^3(J-1)s(15J^2s^2-60Js^2+61s^2+15J^2s-67Js+74s+4J^2-19J+22),\cr
a_4&=&32^4(J-1)s(105J^3s^3-735J^2s^3+1743Js^3-1385s^3+210J^3s^2-1596J^2s^2\cr
&&\quad +4038Js^2-3372s^2+147J^3s-1179J^2s+3114Js-2688s+34J^3-282J^2\cr
&&\quad +765J-674),\cr
a_5&=&32^5(J-1)s(945J^4s^4-10080J^3s^4+40950J^2s^4-74400Js^4+
50521s^4+3150J^4s^3\cr
&&\quad
-35910J^3s^3+153990J^2s^3-292154Js^3+205228s^3+4095J^4s^2-48960J^3s^2\cr
&&\quad +218337J^2s^2-427524Js^2+307860s^2+2370J^4s-29306J^3s+134373J^2s\cr
&&\quad -269085Js+197206s+496J^4-6272J^3+29292J^2-59531J+44134).
\label{e17}
\end{eqnarray}
These moments have some interesting mathematical properties. For example, all
the moments $a_n$, $n\geq 1$, have a factor of $(J-1)s$. Furthermore, in the
residual factor the coefficient of $(Js)^{n-1}$ is $(2n-1)!!$ and the
coefficient of $s^{n-1}$ is the $nth$ Euler number $E_n$.\cite{AS}

The outstanding property of the moments $a_n$ concerns their rapid rate of
growth. This rate of growth can be determined using the fact that there is a
simple relationship between the moments $a_n$ and the coefficients $b_{n-1}$ of
$P_{n-2}(E)$ in the recursion relation (\ref{e6}). Specifically, the Taylor
series
\begin{eqnarray}
f(z)=\sum_{n=0}^{\infty} a_n z^n,
\label{e18}
\end{eqnarray}
whose coefficients are the moments in Eq.~(\ref{e15}), is equivalent to a
continued fraction
\begin{eqnarray}
f(z)=1/(1-b_1z/(1-b_2z/(1-b_3z/(1-\ldots)))),
\label{e19}
\end{eqnarray}
whose coefficients are $b_n$.\cite{BM} Since $b_n$ is a cubic polynomial in $n$
we deduce that the moments $a_n$ grow like $(3n)!$.\cite{BM}

It is unusual to find orthogonal polynomials whose weight functions have
moments
that grow so rapidly. The classical orthogonal polynomials, such as the Hermite
polynomials, typically have moments that grow like $n!$. This is also true of
discrete versions of the classical orthogonal polynomials, such as the Hahn
polynomials.\cite{BMP} The Euler and Bernoulli polynomials are
distinctive\cite{euler} in that
their moments grow like $(2n)!$. However, the polynomials $P_n(E)$ associated
with quasi-exact solvability are of an entirely new type due to the rapid rate
of growth of their moments. Carleman's condition states that when the moments
grow faster than $(2n)!$, the moment problem is not guaranteed to have a unique
solution.\cite{BO} Almost certainly, the weight function $w(E)$ is not unique!
This nonuniqueness corresponds to a kind of gauge invariance that underlies
these quasi-exactly solvable systems. Indeed one may conjecture that the
nonuniqueness of the weight function is related to the Lie algebraic symmetry
of quasi-exact solvability.

\section{ACKNOWLEDGMENTS}
We each thank the U.S. Department of Energy for Financial Support.


\begin{references}
\bibitem{monic} Here, {\sl monic} means that the coefficient of the highest
power of $E$ is one.
\bibitem{domain} Note that the actual domain of integration over $E$ in
(\ref{e12}) is from $-\infty$ to $+\infty$, but depending on the weight
function this domain may be further truncated. Knowing the domain corresponds
to knowing the explicit weight function, which is in general a difficult
problem. However, knowledge of the domain is unnecessary for our algebraic
approach.
\bibitem{euler} The Euler and Bernoulli polynomials differ from the
classical orthogonal polynomials in that they do not satisfy a differential
equation. This is likely to be true of the polynomials $P_n(E)$ discussed in
this paper. Indeed, if the domain of the weight function for a set of
orthogonal polynomials is the whole real line then the polynomials {\sl must}
be the Hermite polynomials.\cite{BF}
\end{references}
\end{document}